\documentclass[preprint,preprintnumbers,amsmath,amssymb,nofootinbib]{revtex4}

\usepackage{graphicx}
\usepackage{amsmath}
\usepackage{amssymb}
\usepackage{graphicx}
\usepackage{dsfont}
\usepackage{overpic}
\usepackage{color}
\usepackage{slashed}
\usepackage{subfigure}
\usepackage{graphicx,textcomp}
\usepackage{natbib}

\allowdisplaybreaks

\begin{document}
\title{On the sigma sigma term}
\author{Peter C.~Bruns}
\affiliation{Institut f\"ur Theoretische Physik, Universit\"at Regensburg, D-93040 Regensburg, Germany}
\date{\today}
\begin{abstract}
We give some estimates for the light-quark mass dependence of the pole position of the sigma ($f_{0}(500)$) resonance in the complex energy plane, 
with the help of a chiral Lagrangian for the resonance field and some input from hadronic models constrained by Chiral Perturbation Theory and elastic unitarity.
We also speculate on the fate of the sigma resonance when the quark masses become unphysically large.
\end{abstract}
\maketitle
\section{Introduction}

There are basically two ways to implement and study resonance phenomena in Chiral Perturbation Theory (ChPT) \cite{Weinberg:1978kz,Gasser:1983yg,Gasser:1987rb}, the low-energy effective field theory of the strong interaction. The first way is the most direct one: a chiral effective Lagrangian is constructed which contains explicit field variables for the particles associated with the resonances. This has lead to the so-called Resonance Chiral Theory \cite{Ecker:1988te,RuizFemenia:2003gw,Cirigliano:2006hb,SanzCillero:2010gy}. The other way is paved by ``Unitarized Chiral Perturbation Theory'' (UChPT) \cite{Dobado:1989qm,Oller:1998hw,Oller:1998zr,Nieves:1999bx,GomezNicola:2001as}, where an infinite string of higher-order terms in the chiral expansion is resummed in some or the other way, to guarantee exact coupled-channel unitarity (in the space of the most relevant particle channels) for a given scattering problem. Here, the resonances enter indirectly: the resummed scattering amplitudes can have poles in the complex energy plane, which are associated with the resonance mass and width. The resonance is said to be ``dynamically generated''. This approach has been very succesful in describing low-energy hadron physics phenomenology, but one should also mention that it has some shortcomings: crossing symmetry and S-matrix analyticity \cite{Eden} are in general not exactly fulfilled (see e.g. \cite{Truong:1991gv,Boglione:1996uz,Nieves:1999bx,Cavalcante:2001eu,Ang:2001bd,Zheng:2003cr,GarciaRecio:2002td}), and there is a non-negligible model dependence \cite{Truong:1991gv,Boglione:1996uz,Caprini:2016uxy} in particular for energies above the low-energy region, and for large quark masses much above the chiral regime where ChPT can be applied (i.~e., problems are expected for $M_{\pi}\gtrsim 350\ldots 400\,\mathrm{MeV}$ \cite{Durr:2014oba}). \\
In \cite{Hanhart:2008mx,Nebreda:2010wv,Pelaez:2010fj,Nebreda:2011di}, the quark mass dependence of the $\sigma$ (or $f_{0}(500)$) (and $\rho$) mass and width has been studied in a UChPT framework. The $\sigma$ resonance is of particular interest for the study of low-energy $\pi\pi$ scattering, a key problem of ChPT. The corresponding pole in the complex energy plane is tightly constrained already from the low-energy $\pi\pi$ interaction given by ChPT and the constraint of elastic unitarity, as pointed out e.g. in Sec.~18 of \cite{Colangelo:2001df}. In a framework where the amplitude is also constrained by partial-wave analyticity and crossing symmetry (the Roy equations for $\pi\pi$ scattering \cite{Roy:1971tc,Ananthanarayan:2000ht}), the $f_{0}(500)$ pole position can be fixed with an impressive accuracy \cite{Caprini:2005zr,GarciaMartin:2011jx}. For a recent comprehensive review on the $\sigma$ resonance, including a historical overview and an extensive list of relevant references, we recommend to consult \cite{Pelaez:2015qba}.\\
It is the aim of this work to establish a close contact between the two ways of describing the $\sigma$ resonance, and to complement the UChPT studies on the quark mass dependence of this resonance by adding another viewpoint to it, in the hope that this may help to further reduce any model dependence, and to contribute to a better understanding of the resonance physics. As an application, we give a first estimate for the leading quark-mass dependence of the mass of the resonance, $m_{\sigma}$ (the phrase ``sigma term'' is only used in loose analogy to the pion-nucleon case - we do not evaluate the scalar form factor of the $\sigma$), and also show some tentative extra\-polations to higher quark masses.\\
This work is organized as follows: In Sec.~\ref{sec:se}, we construct the one-loop approximation to the $\sigma$ self-energy from a resonance chiral Lagrangian, in a similar fashion as we did for the vector mesons in \cite{Bruns:2013tja}. In Sec.~\ref{sec:model}, we exploit the fact that the $\sigma$ resonance is located close to the energy region where (two-loop) ChPT is expected to give reliable results, and that its position is tightly constrained by unitarity and the chiral $\pi\pi$ interaction, to obtain estimates for the most relevant parameters (low-energy constants, or LECs for short) entering the one-loop expression for the self-energy in the chiral limit. In Sec.~\ref{sec:num}, we present and discuss our numerical results for the quark mass dependence of the $\sigma$ pole parameters. The appendix is devoted to a short discussion of the renormalization procedure for theories with explicit resonance fields, in a slightly simplified field-theoretical model.

\newpage

\section{Sigma self-energy}
\label{sec:se}

\begin{figure}[t]
\centering
\includegraphics[width=0.40\textwidth]{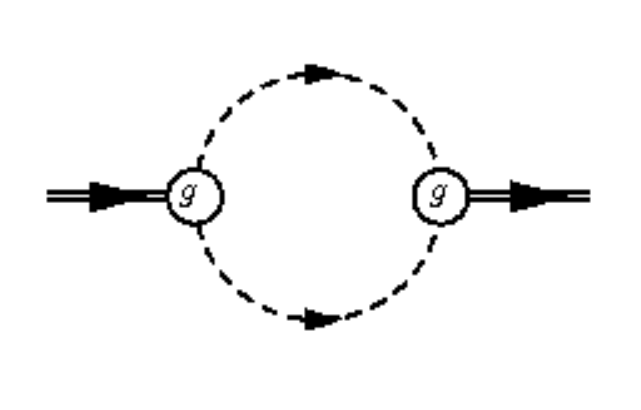}
\caption{One-loop contribution to the $\sigma$ self-energy. Dashed lines: pions, double lines: $\sigma$ resonance.}
\label{fig:sigmaself}
\end{figure}

To begin, we have to write down an effective chiral Lagrangian for the resonance field and its interaction with the pions (the pseudo-Goldstone bosons, PGBs, of spontaneously broken chiral $SU(2)_{L}\times SU(2)_{R}$ symmetry). The leading-order chiral Lagrangian for a massive scalar-isoscalar field $\sigma$ is constructed in \cite{Ecker:1988te} and reads
\begin{equation}
\mathcal{L}_{\sigma} = \frac{1}{2}\partial_{\mu}\sigma \partial^{\mu}\sigma - \frac{1}{2}\mu_{\sigma}^{2}\sigma^{2} + c_{d}^{\sigma}\sigma\langle u_{\mu}u^{\mu}\rangle + c_{m}^{\sigma}\sigma\langle\chi_{+}\rangle\,,\label{eq:Lsigma}
\end{equation}
where the usual chiral-covariant building blocks are used (see also \cite{Ecker:1988te}),
\begin{equation}
u_{\mu} = iu^{\dagger}\left(\partial_{\mu}U\right)u^{\dagger}\,,\quad \chi_{\pm} = 2B\left(u^{\dagger}\mathcal{M}u^{\dagger}\pm u\mathcal{M}u\right)\,,\quad U = \exp\left(\frac{i}{F}\pi^{a}\tau^{a}\right)\,,\quad u=\sqrt{U}\,.
\end{equation}
We have set the external (axial-)vector source fields \cite{Gasser:1983yg} to zero. The isovector pion field is expanded in Pauli matrices $\tau^{a}$, $F$ is the pion decay constant in the chiral limit, $B$ and $c_{d,m}^{\sigma}$ are further low-energy constants, and $\mathcal{M}=\mathrm{diag}(m_{u},\,m_{d})$ is the quark mass matrix (we will work in the isospin limit where $m_{u}=m_{d}=:m_{\ell}$). The quark masses can be expressed through the (squared) pion mass via $M_{\pi}^{2}=2Bm_{\ell}+\mathcal{O}(m_{\ell}^{2}\log m_{\ell})$ \cite{Gasser:1983yg}. We will also introduce counterterms for mass and wave-function renormalization (compare Eq.~(\ref{eq:Lsigmamodel})). These are needed in addition to Eq.~(\ref{eq:Lsigma}) because there is obviously a non-vanishing interaction with pions even in the chiral limit, generating e.g. the width of the sigma resonance (see below). In an effective field theory, an infinite string of higher-order terms with arbitrarily many derivatives is in principle allowed, but we will not need the explicit form of such terms here. The leading quark-mass correction to the counterterm Lagrangian is constructed in analogy to the corresponding term in the pion-nucleon Lagrangian \cite{Gasser:1987rb},
\begin{equation}
\mathcal{L}_{\chi\sigma}^{(2)}:=\frac{c_{1}^{\sigma}}{2}\langle\chi_{+}\rangle\,\sigma^2 + \frac{c_{2}^{\sigma}}{4}\langle u_{\mu}u^{\mu}\rangle\,\sigma^2\,,\quad \mathcal{L}_{\chi\sigma}^{(4)}:= -\frac{e_{1}^{\sigma}}{32}\langle\chi_{+}\rangle^2\sigma^2+\ldots\,.
\end{equation}
Chiral Lagrangians involving resonances (instead of particles like pions or nucleons, which are stable under the strong interaction) like the ones given above have to be applied and interpreted with care. First, quantum (field) theory is based on measurements. For a broad (short-lived) resonance like the $\sigma$, there seems to be no obvious or natural concept of a ``$\sigma$ state'', let alone multiparticle $\sigma$ states, and consequently, one might also have doubts to use a $\sigma$ field\footnote{One could, however, resort to a formalism as proposed in \cite{Berggren:1968zz,Garcia-Calderon:1976omn}, at least for narrow resonances.}. Second, the renormalization process is non-standard for resonances (see also App.~\ref{app:rconds}). And third, since the large width of the $\sigma$ is generated in the perturbative loop expansion in such a framework, it is possible that the resonance couples so strongly to other states that the perturbative expansion does not converge. \\
Therefore, it is clear that the present study has some exploratory character. We see the resonance Lagrangian as a convenient tool to parameterize and describe the phenomenon observed as the resonance, and the $\sigma$ field as a (largely arbitrary, up to the quantum numbers) integration variable in the path integral which is used in the description of the relevant observations, not implying any assumptions about the ``nature'' (quark content, etc.) of the resonance. The theory at hand can be considered as a natural generalization of the framework designed for (nearly) stable particles, respecting all known symmetries relevant at low energies, but also showing some unusual features which will be encountered in the present work (compare also \cite{Achasov:2004uq} for a theoretical study of general properties of the $\sigma$ propagator). \\
We add a remark on the power counting. In the usual ChPT framework \cite{Weinberg:1978kz,Gasser:1983yg}, PGB masses, momenta and energies are counted as being ``small of order $\mathcal{O}(p)$'', since these energies are small compared to a typical hadronic scale $\sim 1\,\mathrm{GeV}\sim 4\pi F$. Even though the mass and width of the $\sigma$ do not vanish in the chiral limit, the resonance energy region is not far away from the region where the low-energy expansion properly works. On a practical level, an expansion in the energy over $4\pi F$ could still be reasonably effective. For the application of the self-energy intended here, it will turn out that the expansion in small energies is of minor importance, since we are mainly interested in quark mass corrections to the pole position of the resonance in the chiral limit. We will rearrange our expression for the self-energy in a way that is convenient for this purpose, similar to our work on vector mesons in \cite{Bruns:2013tja}, but without some of the approximations made therein. We restrict our application of the chiral resonance Lagrangian to the one-loop level, which is $\mathcal{O}(p^4)$ in the usual low-energy counting, but we will mainly be interested in the leading quark-mass dependence (of $\mathcal{O}(M_{\pi}^{2})$) of the resonance position here, and will therefore neglect some higher-order corrections of $\mathcal{O}(M_{\pi}^{4})$ in our final numerical estimates, which cannot be fixed without additional input.\\
The most prominent contribution to the $\sigma$ self-energy is given by the one-loop graph of Fig.~\ref{fig:sigmaself}, which describes the coupling of the resonance to the $\pi\pi$ continuum states, with a strength given by some coupling parameter $g$. From the vertex rules of the Lagrangian (\ref{eq:Lsigma}) it is straightforward to compute the contribution to the self-energy $\Pi_{\sigma}(s)$ due to this Feynman graph,
\begin{equation}
\Pi_{\sigma\mid\pi\pi}(s) = \frac{24}{F^4}\biggl(\left[\frac{c^{\sigma}_{d}}{2}(s-2M_{\pi}^{2})+c^{\sigma}_{m}M_{\pi}^{2}\right]^2I_{\pi\pi}(s)-\left((c^{\sigma}_{d})^2(s-4M_{\pi}^{2})+4c^{\sigma}_{m}c^{\sigma}_{d}M_{\pi}^{2}\right)\frac{I_{\pi}}{2}\biggr)\,.
\end{equation}
Here $s$ is the squared four-momentum of the resonance. In dimensional regularization, the loop integrals entering here are given by
\begin{displaymath}
I_{\pi}:=\int\frac{d^{d}l}{(2\pi)^{d}}\frac{i}{l^2-M_{\pi}^2} = 2M_{\pi}^{2}\bar{\lambda} + \frac{M_{\pi}^{2}}{16\pi^2}\log\left(\frac{M_{\pi}^{2}}{\mu^2}\right) + \mathcal{O}(4-d)\,,
\end{displaymath}
where $\bar{\lambda}$ contains the pole in $d-4$ and depends on the regularization scale $\mu$ such that the $\mu$-dependence of the logarithm is cancelled, while $I_{\pi\pi}$ can be obtained from Eq.~(\ref{eq:disp1}) with $M_{\phi}\rightarrow M_{\pi}$. - Let us first discuss the representation of the self-energy in the chiral limit where $M_{\pi}\rightarrow 0$. In this limit (indicated by a superscript $\circ$), $\Pi_{\sigma\mid\pi\pi}$ reduces to
\begin{equation}
\overset{\circ}{\Pi}_{\sigma\mid\pi\pi}(s) = \frac{3(c^{\sigma}_{d})^2s^2}{8\pi^2F^4}\left(32\pi^2\bar{\lambda}-1-\log\left(-\frac{\mu^2}{s}\right)\right)\,.
\end{equation}
Of course, in the chiral limit, decays into any higher (even) number of PGBs are kinematically allowed, but the corresponding contributions to the self-energy are still suppressed in the low-energy counting, and thus the use of a one-loop approximation can still be meaningful. 

The inverse propagator of the resonance at the one-loop level is then of the form
\begin{equation}
(\overset{\circ}{D}_{\sigma})^{-1}(s) = s-\overset{\circ}{\mu}_{\sigma}^{\raisebox{-0.15cm}{\scriptsize{2}}}+\frac{3\overset{\circ}{g}_{\sigma}^{\raisebox{-0.15cm}{\scriptsize{\,2}}}}{32\pi^2}s^2\log\left(-\frac{\mu^2}{s}\right) + \mathrm{polynomial}(s)\,,\quad \overset{\circ}{g}_{\sigma}:=\frac{2c_{d}^{\sigma}}{F^2}\,.
\end{equation}
The polynomial is due to the analytic piece in the loop function and counterterms from the Lagrangian. Applyling the low-energy counting, the polynomial would be restricted to second order in $s$. But the order is not really relevant here, since we are only interested in the behavior of this function close to the pole position of the propagator on the second Riemann sheet in the variable $s$, written as
\begin{equation}\label{eq:spolchlim}
\overset{\circ}{s}_{\sigma}\equiv \overset{\circ}{\mu}_{\sigma}^{\raisebox{-0.15cm}{\scriptsize{2}}}-i|\overset{\circ}{\mu}_{\sigma}|\overset{\circ}{\gamma}_{\sigma}\equiv \left(\overset{\circ}{m}_{\sigma}-\frac{i}{2}\overset{\circ}{\Gamma}_{\sigma}\right)^2\,,
\end{equation}
so that, for $s$ sufficiently close to the pole,
\begin{equation}\label{eq:Dsigma}
(\overset{\circ}{D}_{\sigma})^{-1}(s) = s-\overset{\circ}{\mu}_{\sigma}^{\raisebox{-0.15cm}{\scriptsize{2}}}+\frac{3\overset{\circ}{g}_{\sigma}^{\raisebox{-0.15cm}{\scriptsize{\,2}}}}{32\pi^2}s^2\log\left(-\frac{\mu^2}{s}\right) + \kappa_{0}+\kappa_{1}(s-\overset{\circ}{\mu}_{\sigma}^{\raisebox{-0.15cm}{\scriptsize{2}}})+\kappa_{2}(s-\overset{\circ}{\mu}_{\sigma}^{\raisebox{-0.15cm}{\scriptsize{2}}})^2+\ldots\,.
\end{equation}
Terms of quadratic and higher order in $(s-\overset{\circ}{\mu}_{\sigma}^{\raisebox{-0.15cm}{\scriptsize{2}}})$ contain higher powers of the width for $s\rightarrow\overset{\circ}{s}_{\sigma}$ and are beyond the one-loop order, so we can drop them for our purposes.
 In the following, we will fix $\mu=1\,\mathrm{GeV}$, and sometimes write the logarithm in the loop function simply as $\log\left(-\frac{1}{s}\right)$ with the unit being understood. Of course, the numerical values of the coefficients $\kappa_{i}$ will in general depend on the choice of $\mu$. The logarithm has branch points at zero and infinity, and we choose the pertinent branch cut along the positive real $s$-axis. The first Riemann sheet ($I$) is the one where the $\log$ is real on the negative $s$-axis. On the second Riemann sheet ($II$), $\log\left(-\frac{1}{s}\right)|_{II}=\log\left(-\frac{1}{s}\right)|_{I}+2\pi i$. The $\kappa_{0}$-term is a mass counterterm, which has to be adjusted such that $\overset{\circ}{\mu}_{\sigma}^{\raisebox{-0.15cm}{\scriptsize{2}}}$ is indeed the real part of the pole position, while the $\kappa_{1}$-term is essentially a wave-function normalization counterterm. Of course, the complex coefficients $\kappa_{i}$ must be such that the counterterm polynomial is real for real $s$. To proceed further, we must establish a connection to the $\pi\pi$ scattering amplitude (in the chiral limit). First, we note that, on tree level, the exchange of a $\sigma$ resonance in the $s-$channel gives a pole-graph contribution 
\begin{equation}
(t_{0}^{0}(s))_{\sigma-\mathrm{ex.}}^{(\mathrm{tree})} = -\frac{3}{32\pi}\frac{\overset{\circ}{g}_{\sigma}^{\raisebox{-0.15cm}{\scriptsize{\,2}}}s^2}{s-\overset{\circ}{\mu}_{\sigma}^{\raisebox{-0.15cm}{\scriptsize{2}}}}
\end{equation}
to the isospin $I=0$ s-wave ($J=0$) partial-wave $\pi\pi$ scattering amplitude $t^{I=0}_{J=0}(s)$ (defined as in \cite{Gasser:1983yg}). This violates elastic unitarity and is certainly not a good approximation for a broad resonance. But, taking the ``dressing'' of the resonance into account, we can write an improved amplitude
\begin{equation}\label{eq:t00sigmaex}
(t_{0}^{0}(s))_{\sigma-\mathrm{ex.}} = -\frac{3\overset{\circ}{g}_{\sigma}^{\raisebox{-0.15cm}{\scriptsize{\,2}}}}{32\pi}s^2\overset{\circ}{D}_{\sigma}(s)\,,
\end{equation}
which satisfies the constraint of elastic unitarity (stated in Eq.~(\ref{eq:unichlim}) below) and should be a good approximation to the full partial-wave scattering amplitude close to the resonance pole, provided that the pole position and coupling are properly adjusted. Assuming that we know the pole position and (complex) residue $\overset{\circ}{R}_{\sigma}$ of the full scattering amplitude,
\begin{equation}\label{eq:t00atpole}
t_{0}^{0}(s\rightarrow \overset{\circ}{s}_{\sigma}) = \frac{\overset{\circ}{R}_{\sigma}}{s-\overset{\circ}{s}_{\sigma}} \equiv -\frac{3\overset{\circ}{g}_{\sigma}^{\raisebox{-0.15cm}{\scriptsize{\,2}}}}{32\pi}\overset{\circ}{s}_{\sigma}^{\raisebox{-0.15cm}{\scriptsize{\,2}}}\frac{\mathcal{Z}_{1}+i\mathcal{Z}_{2}}{s-\overset{\circ}{s}_{\sigma}}\,,\quad \mathcal{Z}_{1,2}\in\mathds{R}\,,
\end{equation}
we can try to adjust our free parameters in (\ref{eq:Dsigma}) and (\ref{eq:t00sigmaex}) to obtain the required approximation. We will use the renormalization conditions that the mass in the chiral limit is given by $\overset{\circ}{\mu}_{\sigma}$, and that the real part of the derivative of the self-energy vanishes, $\mathrm{Re}\,\overset{\circ}{\Pi'}_{\sigma}(\overset{\circ}{s}_{\sigma})=0$ (which approximately fixes $\mathcal{Z}_{1}=1$, compare App.~\ref{app:rconds}). We find 
\begin{equation}
\kappa_{0} = -\frac{3\overset{\circ}{g}_{\sigma}^{\raisebox{-0.15cm}{\scriptsize{\,2}}}}{32\pi^2}\mathrm{Re}\left[\overset{\circ}{s}_{\sigma}^{\raisebox{-0.15cm}{\scriptsize{\,2}}}\log\left(-\frac{\mu^2}{\overset{\circ}{s}_{\sigma}}\right)\biggr|_{II}\right]\,,\quad 
\kappa_{1} = -\frac{3\overset{\circ}{g}_{\sigma}^{\raisebox{-0.15cm}{\scriptsize{\,2}}}}{32\pi^2}\mathrm{Re}\left[\overset{\circ}{s}_{\sigma}\left(2\log\left(-\frac{\mu^2}{\overset{\circ}{s}_{\sigma}}\right)\biggr|_{II} -1\right)\right]\,,\label{eq:fixkappa}
\end{equation}
\begin{equation}\label{eq:fixg}
\frac{3\overset{\circ}{g}_{\sigma}^{\raisebox{-0.15cm}{\scriptsize{\,2}}}}{32\pi^2} = \frac{|\overset{\circ}{\mu}_{\sigma}|\overset{\circ}{\gamma}_{\sigma}}{|\overset{\circ}{\mu}_{\sigma}|\overset{\circ}{\gamma}_{\sigma}\mathrm{Re}\left[\overset{\circ}{s}_{\sigma}\left(2\log\left(-\frac{\mu^2}{\overset{\circ}{s}_{\sigma}}\right)\biggr|_{II} -1\right)\right] + \mathrm{Im}\left[\overset{\circ}{s}_{\sigma}^{\raisebox{-0.15cm}{\scriptsize{\,2}}}\log\left(-\frac{\mu^2}{\overset{\circ}{s}_{\sigma}}\right)\biggr|_{II}\right]}\,.
\end{equation}
As usual in ChPT, the quark masses, and therefore the pion masses, are treated as an additional external perturbation. The pole position is shifted by this perturbation, to
\begin{equation}\label{eq:spolphys}
s_{\sigma}\equiv \mu_{\sigma}^{2}-i|\mu_{\sigma}|\gamma_{\sigma}\equiv \left(m_{\sigma}-\frac{i}{2}\Gamma_{\sigma}\right)^2\,,
\end{equation}
and the one-loop approximation to the inverse propagator is of the form
\begin{eqnarray}
(D_{\sigma})^{-1}(s) &=& s-\overset{\circ}{\mu}_{\sigma}^{\raisebox{-0.15cm}{\scriptsize{2}}} + \kappa_{0}+4c_{1}^{\sigma}M_{\pi}^{2}\left(1-\frac{3M_{\pi}^{2}}{32\pi^2F^2}\log\left(\frac{M_{\pi}^{2}}{\mu^2}\right)\right) + \frac{3M_{\pi}^{4}c_{2}^{\sigma}}{16\pi^2F^2}\log\left(\frac{M_{\pi}^{2}}{\mu^2}\right) - e_{1}^{\sigma}M_{\pi}^{4} \nonumber\\
 &+& \left(\kappa_{1}+4\kappa_{1}'M_{\pi}^{2}\right)(s-\overset{\circ}{\mu}_{\sigma}^{\raisebox{-0.15cm}{\scriptsize{2}}}) +\frac{3M_{\pi}^{2}}{4\pi^2F^4}\left((c^{\sigma}_{d})^2(s-4M_{\pi}^{2})+4c^{\sigma}_{m}c^{\sigma}_{d}M_{\pi}^{2}\right)\log\left(\frac{M_{\pi}^{2}}{\mu^2}\right) \label{eq:DsigmaFull}\\
 &+& \frac{3}{2\pi^2F^4}\left[\frac{c^{\sigma}_{d}}{2}(s-2M_{\pi}^{2})+c^{\sigma}_{m}M_{\pi}^{2}\right]^2\left(2\sigma(s)\,\mathrm{artanh}\left(-\frac{1}{\sigma(s)}\right)-\log\left(\frac{M_{\pi}^{2}}{\mu^2}\right)\right)\,,\nonumber
\end{eqnarray}
where we have absorbed some analytic pieces of the loop integrals in the (renormalized) LECs (also, the leading corrections to the mass formula $M_{\pi}^{2}=2Bm_{\ell}$ have been tacitly absorbed in $c_{2}^{\sigma},e_{1}^{\sigma}$). The new LEC $\kappa_{1}'$ is due to a quark-mass correction to the wave-function renormalization constant. On the unphysical sheet, $\sigma(s)\mathrm{artanh}\left(-1/\sigma(s)\right)_{II}=\sigma(s)\mathrm{artanh}\left(-1/\sigma(s)\right)_{I}+i\pi\sigma(s)$, $\sigma(s)=\sqrt{1-(4M_{\pi}^{2}/s)}\,$. Eq.~(\ref{eq:DsigmaFull}) is a main result of this work: if the LECs are known, the equation $(D_{\sigma})^{-1}(s_{\sigma})\overset{!}{=}0$ determines the complex pole position $s_{\sigma}$ of the $\sigma$ for any prescribed value of $M_{\pi}$. In the next section, we want to obtain estimates for the most important parameters entering the above equation.

\newpage

\section{Chiral unitary model in the chiral limit}
\label{sec:model}

The following partial-wave amplitudes $t^{I}_{J}(s)$ in the chiral limit satisfy the requirement of elastic two-particle unitarity,
\begin{equation}\label{eq:unichlim}
\mathrm{Im}\left[(t^{I}_{J}(s))^{-1}\right] = -1\,,\quad\mathrm{for}\quad s>0\,,
\end{equation}
and agree with the known chiral expansion at the two-loop level \cite{Bijnens:1995yn,Hannah:1997ux}\,,
\begin{tiny}
\begin{align*}
t^{0}_{0}(s)\, & = & \hspace{-0.4cm}\frac{s}{16\pi F^2}\biggl[1 +& \frac{s}{(4\pi F)^2}\biggl(r_{0}^{0}s - \frac{17+64\pi^2(11l_{1}^{r}+7l_{2}^{r})}{12} - \log\left(\frac{1}{s}\right)\left(\frac{7}{18}-\frac{s}{1296(4\pi F)^2}\left(1255+960\pi^2(82 l_{1}^{r}+29 l_{2}^{r})\right)\right)+\frac{175s}{324(4\pi F)^2}\log^{2}\left(\frac{1}{s}\right)\biggr) \\ & & -&\frac{s}{(4\pi F)^2}\log\left(-\frac{1}{s}\right)\biggr]^{-1}\,,\\
t^{1}_{1}(s)\, & = & \hspace{-0.4cm}\frac{s}{96\pi F^2}\biggl[1 +& \frac{s}{(4\pi F)^2}\biggl(r_{1}^{1}s + 16\pi^2(2l_{1}^{r}-l_{2}^{r})-\frac{1}{9} + \log\left(\frac{1}{s}\right)\left(\frac{1}{6}-\frac{s}{240(4\pi F)^2}\left(737 +576\pi^2(34 l_{1}^{r} + 33 l_{2}^{r})\right)\right) - \frac{5s}{4(4\pi F)^2}\log^{2}\left(\frac{1}{s}\right)\biggr) \\ & & -&\frac{s}{6(4\pi F)^2}\log\left(-\frac{1}{s}\right)\biggr]^{-1}\,,\\
t^{2}_{0}(s)\, & = & \hspace{-0.4cm}\,\,-\frac{s}{32\pi F^2}\biggl[1 +& \frac{s}{(4\pi F)^2}\biggl(r_{0}^{2}s + \frac{51+768\pi^2(l_{1}^{r}+2l_{2}^{r})}{36} + \log\left(\frac{1}{s}\right)\left(\frac{11}{18}+\frac{s}{1296(4\pi F)^2}\left(883+192\pi^2(193 l_{2}^{r}-46 l_{1}^{r})\right)\right) + \frac{85s}{324(4\pi F)^2}\log^{2}\left(\frac{1}{s}\right)\biggr) \\ & & +&\frac{s}{2(4\pi F)^2}\log\left(-\frac{1}{s}\right)\biggr]^{-1}\,.
\end{align*}
\end{tiny}
The unknown constants $r^{I}_{J}$ appear at two-loop order and contain the renormalized two-loop LECs $r_{i}(\mu=1\,\mathrm{GeV})$. The one-loop LECs $l_{1,2}^{r}$ are (roughly) known; for definiteness, we will employ the values (and errors) of \cite{Colangelo:2001df}, $\bar{l}_{1}=-0.4\pm 0.6$, $\bar{l}_{2}=4.3\pm 0.1$, which translates to $l_{1}^{r}(\mu=1\,\mathrm{GeV})=(-4.5\pm 0.6)\cdot 10^{-3}$ and $l_{2}^{r}(\mu=1\,\mathrm{GeV})=(0.8\pm 0.2)\cdot 10^{-3}$. For a discussion of these (and other) LECs relevant for $\pi\pi$ scattering we refer to \cite{Nebreda:2012ve}. The pion decay constant in the chiral limit will be taken as $F=86\,\mathrm{MeV}$ \cite{Aoki:2016frl}.

\begin{figure}[h]
\centering
\subfigure[]{\includegraphics[width=0.38\textwidth]{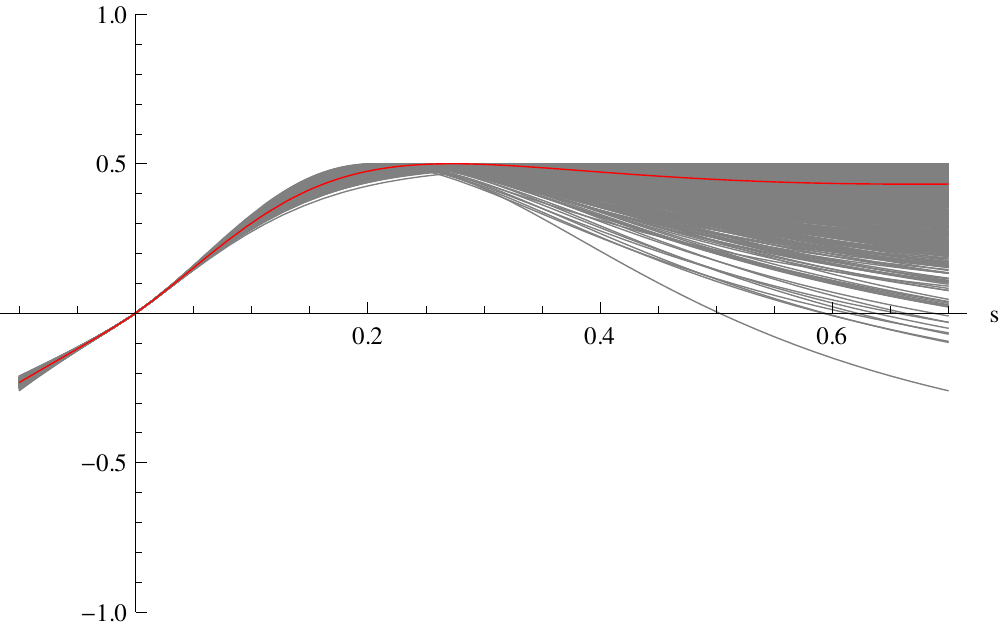}} \\
\subfigure[]{\includegraphics[width=0.31\textwidth]{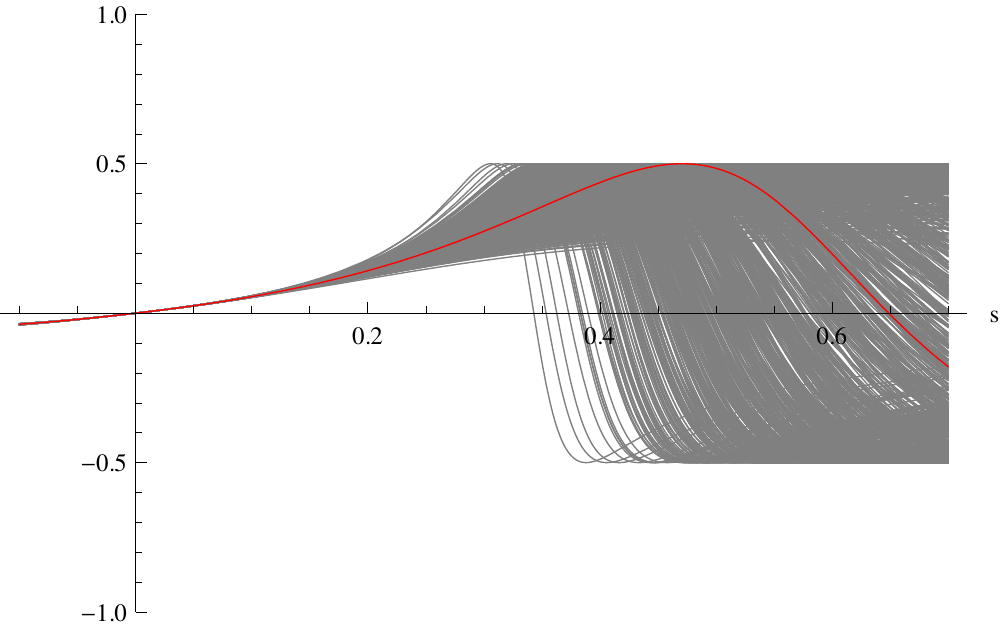}}\hspace{1cm}
\subfigure[]{\includegraphics[width=0.31\textwidth]{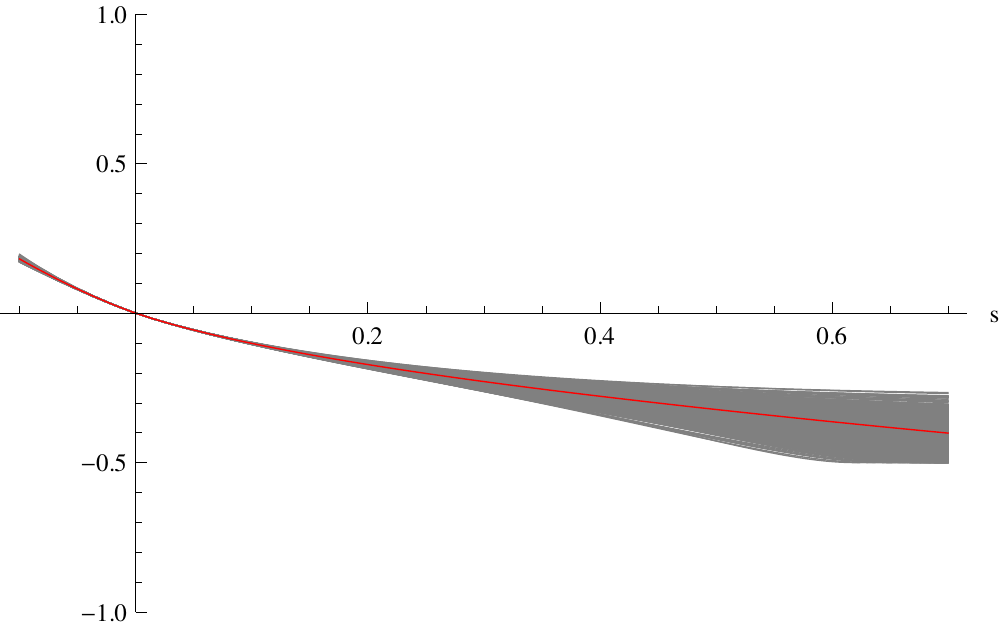}}
\caption{$\mathrm{Re}\,t^{I}_{J}(s)$, over $s$ in $\mathrm{GeV}^2$ (a): IJ=00, (b): IJ=11, (c) IJ=20.}
\label{fig:retIJ}
\end{figure}

There is a resonance pole on the second Riemann sheet of the above model amplitude for $t_{0}^{0}(s)$. For given values of $l_{1}^{r}$, $l_{2}^{r}$ and $r_{0}^{0}$, we can extract its position with the help of $\mathrm{mathematica}\textsuperscript{\textregistered}$ routines. To estimate the uncertainty, we generate $\sim 10^4$ random number sets for $(l_{1}^{r},\,l_{2}^{r},\,r_{0}^{0})$, normally distributed around their central values, with a standard deviation of the corresponding error. Values for $r_{0}^{0}$ are taken around $0$ with a deviation of $1\,\mathrm{GeV}^{-2}$, which is the order of magnitude of the numerically fixed terms in the two-loop calculation entering the combination $r_{0}^{0}$. This range is quite generous and might also cover some systematic error due to the model-dependence involved in the choice of the unitarization procedure. We show the result of this enterprise (for the ``predicted'' $\mathrm{Re}\,t_{0}^{0}$) in Fig.~\ref{fig:retIJ}(a). The analogous plots for  $\mathrm{Re}\,t_{1}^{1}$ and  $\mathrm{Re}\,t_{0}^{2}$ are shown in Figs.~\ref{fig:retIJ}(b,c). The red lines show the result for the central values of the input parameters, the curves due to all other parameter configurations form the gray bands. The uncertainty of the parameterization seems to be relatively moderate in the $\sigma$ resonance region, while the situation appears to be worse for the $\rho$. For the pole position of the $\sigma$ in the chiral limit, $\overset{\circ}{s}_{\sigma}=(\overset{\circ}{m}_{\sigma}-\frac{i}{2}\overset{\circ}{\Gamma}_{\sigma})^2$, and the residue at the pole, $\overset{\circ}{R}_{\sigma}$, we find
\begin{eqnarray}
\overset{\circ}{m}_{\sigma} &=& (395\,\pm\,17)\,\mathrm{MeV}\,,\qquad \overset{\circ}{\Gamma}_{\sigma} = (630\,\pm\,135)\,\mathrm{MeV}\,,\label{est:polchlim}\\
\overset{\circ}{R}_{\sigma} &=& (0.183\,\pm\,0.036)\,\mathrm{GeV}^2 + i\,(0.057\,\pm\,0.028)\,\mathrm{GeV}^2\,.\label{est:reschlim}
\end{eqnarray}
The uncertainty in the mass is surprisingly small and one might suspect that it is underestimated. Nevertheless, we will adopt the values given above for our estimates of the quark mass dependence. For the record, we note that we find the pole position of the lowest resonance in $t_{1}^{1}$ (the $\rho$) at
\begin{eqnarray}
\overset{\circ}{m}_{\rho} &=& (711\,\pm\,45)\,\mathrm{MeV}\,,\qquad \overset{\circ}{\Gamma}_{\rho} = (333\,\pm\,179)\,\mathrm{MeV}\,,\\
\overset{\circ}{R}_{\rho} &=& (-0.039\,\pm\,0.089)\,\mathrm{GeV}^2 + i\,(0.076\,\pm\,0.100)\,\mathrm{GeV}^2\,.
\end{eqnarray}
We point out that the numbers given above, and the curves in Fig.~\ref{fig:retIJ}, are {\em not\,} due to a fit to data - the data enter only indirectly through the numerical values of $F,\,l_{1}^{r},\,l_{2}^{r}$ (and to a lesser extent ($\mathcal{O}(p^8)$) also through the fixed $\mu=1\,\mathrm{GeV}$). Taking for granted the result of Ref.~\cite{Caprini:2005zr} (compare also \cite{Pelaez:2015qba}),
\begin{equation}\label{est:polphys}
m_{\sigma}^{\mathrm{phys}} = 441^{+16}_{-8}\,\mathrm{MeV}\,,\qquad \Gamma_{\sigma}^{\mathrm{phys}} = 544^{+18}_{-25}\,\mathrm{MeV}\,,
\end{equation}
we obtain a first rough estimate of the ``$\sigma$ sigma term'', $m_{\ell}\frac{\partial m_{\sigma}}{\partial m_{\ell}}\approx m_{\sigma}^{\mathrm{phys}}-\overset{\circ}{m}_{\sigma}\sim 45\,\mathrm{MeV}\,$.\\


\section{Numerical analysis and discussion}
\label{sec:num}

Let us now use the estimates obtained in the previous section to study the quark mass dependence of the resonance pole position. Since we cannot expect more than first rough estimates without analyzing precise lattice data, it makes sense to set the $\mathcal{O}(M_{\pi}^{4})$ parameters $e_{1}^{\sigma},\,c_{2}^{\sigma}$ in Eq.~(\ref{eq:DsigmaFull}) to zero for the moment. Our numerical strategy will be the same as in the previous section: We generate a large set of random numbers for $(\overset{\circ}{\mu}_{\sigma},\overset{\circ}{\gamma}_{\sigma},\mu_{\sigma},\gamma_{\sigma},c_{m}^{\sigma})$ (see Eqs.~(\ref{eq:spolchlim}),~(\ref{eq:spolphys})) with central values and errors as specified in Eqs.~(\ref{est:polchlim}), (\ref{est:polphys}) (while $c_{m}^{\sigma}$ is varied in the range $-50\ldots +50\,\mathrm{MeV}$, which is the expected range for this coupling, compare e.g. Secs.~4, 5 of \cite{Ecker:1988te}), and determine the LECs $c_{1}^{\sigma}$ and $\kappa_{1}'$, for every set, from the equation $(D_{\sigma})^{-1}(s_{\sigma})\overset{!}{=}0$ for $M_{\pi}=139\,\mathrm{MeV}$.\\
As a first result, we obtain (via Eq.~(\ref{eq:fixg})) an estimate for the coupling in the chiral limit,
\begin{equation}\label{est:g}
\frac{3\overset{\circ}{g}_{\sigma}^{\raisebox{-0.15cm}{\scriptsize{\,2}}}}{32\pi^2} = (0.886 \pm 0.170)\,\mathrm{GeV}^{-2}\,,
\end{equation}
or $|c_{d}^{\sigma}|\approx 36\,\mathrm{MeV}$. This coupling is quite large and we have to expect that higher-order corrections could be sizeable. To provide a test of the validity of our approach, we note that the residue of our model amplitude in the chiral limit, Eq.~(\ref{eq:t00sigmaex}), is also fixed via Eqs.~(\ref{est:g}), (\ref{eq:fixkappa}), and results in $\overset{\circ}{R}_{\sigma}\approx (0.169 + 0.067i)\,\mathrm{GeV}^{2}$ (so that $\mathcal{Z}_{1}\approx 1\,$, $\mathcal{Z}_{2}\approx -0.04$ in Eq.~(\ref{eq:t00atpole})), which can be compared with Eq.~(\ref{est:reschlim}). It is clear that the given residue in the chiral limit can not be reproduced exactly, because we work only at one-loop accuracy, {\em and\,} because vertex corrections are missing in the simplistic model used in Eq.~(\ref{eq:t00sigmaex}) (see also App.~\ref{app:rconds}). The impact of these deficiencies seems to be moderate, however. \\
The values and mean errors for the LECs determined from the procedure described above are
\begin{equation}\label{est:c1sigma}
c_{1}^{\sigma} = 0.35 \pm 1.36\,,\qquad \kappa_{1}' = (0.53 \pm 3.39)\,\mathrm{GeV}^{-2}\,.
\end{equation}
As already anticipated, the uncertainties are relatively large. This is reflected by the estimated quark mass dependence of the mass $m_{\sigma}$ and the width $\Gamma_{\sigma}$ for small $M_{\pi}$: Inserting an ansatz
\begin{equation}\label{eq:leadingqm}
m_{\sigma}  = \overset{\circ}{m}_{\sigma} + a_{m}M_{\pi}^{2}\log M_{\pi}^{2} + b_{m}M_{\pi}^{2}\,,\quad \Gamma_{\sigma}  = \overset{\circ}{\Gamma}_{\sigma} + a_{\Gamma}M_{\pi}^{2}\log M_{\pi}^{2} + b_{\Gamma}M_{\pi}^{2}\,
\end{equation}
in Eq.~(\ref{eq:DsigmaFull}), expanding everything to order $M_{\pi}^{2}$, and solving this truncated version of the equation $(D_{\sigma})^{-1}(s_{\sigma})\overset{!}{=}0$ for all generated parameter configurations, we obtain
\begin{equation}\label{est:ab}
a_{m}=a_{\Gamma}=0\,,\quad b_{m}= (1.95 \pm 1.65)\,\mathrm{GeV}^{-1}\,,\quad b_{\Gamma}= (-6.70 \pm 6.95)\,\mathrm{GeV}^{-1}\,,
\end{equation}
resulting in the estimate $\sim 2M_{\pi,\mathrm{phys}}^{2}/\mathrm{GeV}$, or $(38\pm 32)\,\mathrm{MeV}$ for the ``sigma sigma term''. Of course, this result, and the numbers in Eq.~(\ref{est:ab}) do not provide much more than a consistency check between the estimates in Sec.~\ref{sec:model}, the framework outlined in Sec.~\ref{sec:se} and the methods used here. \\

\begin{figure}[h]
\centering
\subfigure[]{\includegraphics[width=0.49\textwidth]{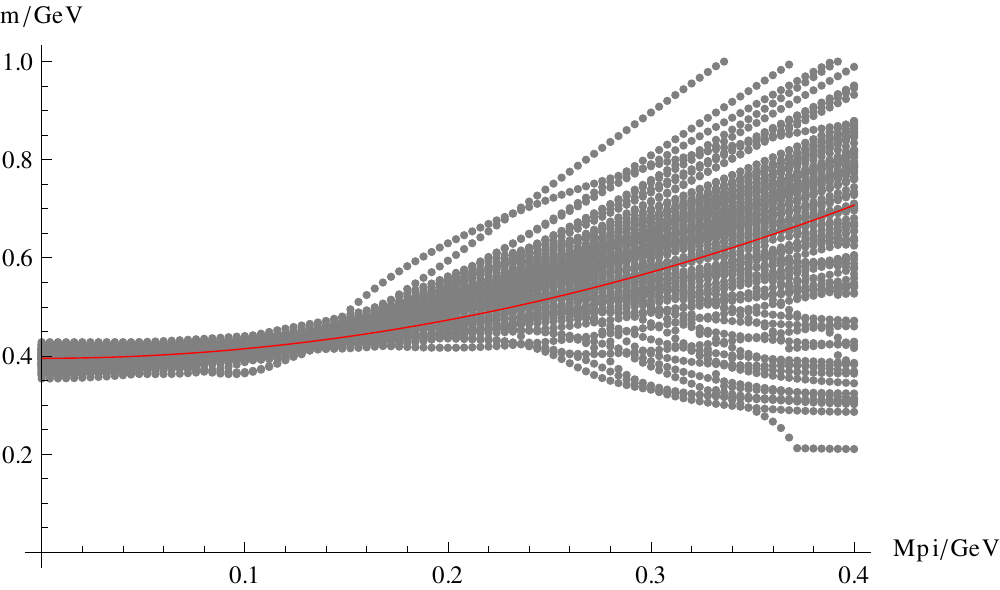}} 
\subfigure[]{\includegraphics[width=0.49\textwidth]{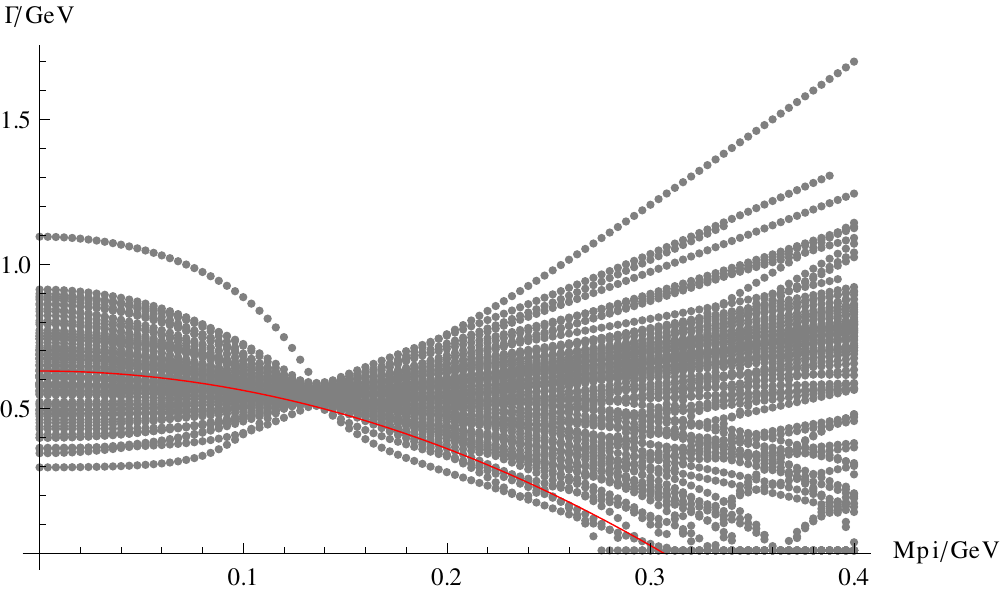}}
\caption{Mass (a) and width (b) of the $\sigma$ resonance for $10^2$ solutions of $D_{\sigma}^{-1}(s_{\sigma})\overset{!}{=}0$.}
\label{fig:spole_of_Mpi}
\end{figure}

It is probably bold to extrapolate our results so far to unphysically large pion masses. But in order to see what the present formalism is in principle capable of, we show in Fig.~\ref{fig:spole_of_Mpi} the pion mass dependence of the mass $m_{\sigma}$ and width $\Gamma_{\sigma}$, for a hundred of our solutions for the zeroes of Eq.~(\ref{eq:DsigmaFull}). The red curves give the leading quark mass dependence of Eq.~(\ref{eq:leadingqm}), for the central values of Eqs.~(\ref{est:polchlim}), (\ref{est:ab}). Note that our analysis is not complete at $\mathcal{O}(p^4)$, because we have set the parameters $c_{2}^{\sigma}$, $e_{1}^{\sigma}$ to zero by hand, so that the resulting curves can only give a first impression of the full one-loop solution. While the mass follows the description given by the leading quark-mass dependence up to rather high pion masses, the uncertainty in the width is large beyond the physical point. But at least it seems fair to say that the $\sigma$ can become a bound state not much below $M_{\pi}\sim 300\,\mathrm{MeV}$ (if at all). This is in accord with a recent lattice study \cite{Briceno:2016mjc}. More definite conclusions can only be drawn if the present formalism is applied to an analysis of precise lattice data. This is a natural next step in the development of this study.\\

\acknowledgments{I thank Maxim Mai, Andreas Sch\"afer and Philipp Wein for discussions on the manuscript. This work was supported by the Deutsche Forschungsgemeinschaft SFB/Transregio 55.}

\newpage

\begin{appendix}

\section{Renormalization conditions for resonances}
\label{app:rconds}
\def\theequation{\Alph{section}.\arabic{equation}}
\setcounter{equation}{0}

\begin{figure}[t]
\centering
\subfigure[]{\includegraphics[width=0.54\textwidth]{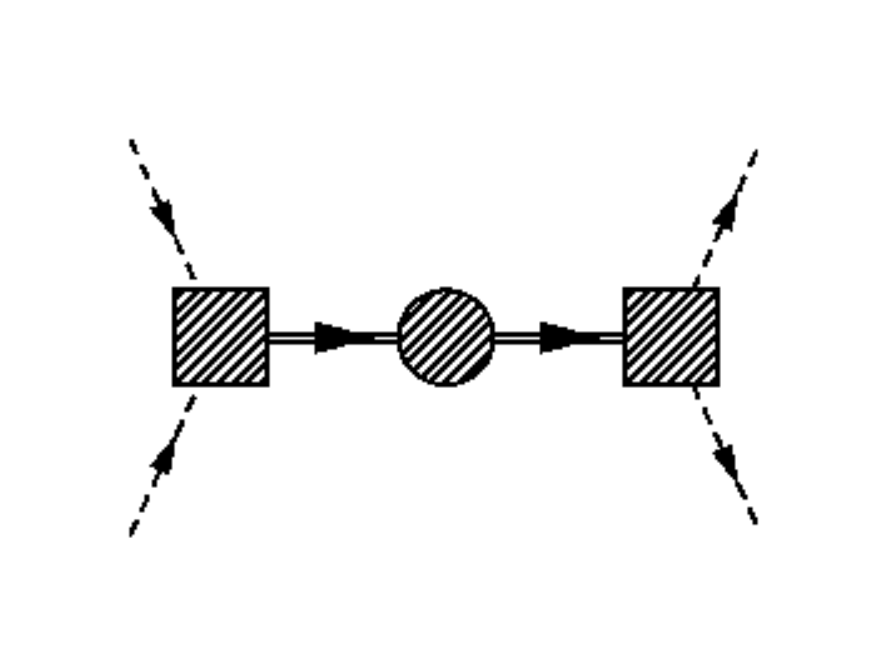}}\hspace{1cm}
\subfigure[]{\includegraphics[width=0.34\textwidth]{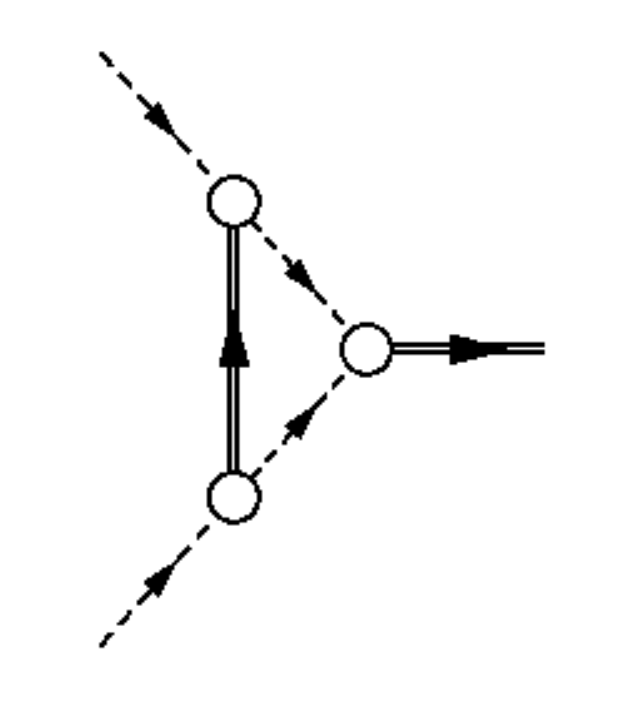}}
\caption{(a) General structure of the resonance pole contribution, (b) a one-loop vertex correction. Dashed lines: $\phi$ particles, double lines: $\sigma$ resonance. The squares in (a) contain the vertex corrections, the blob on the resonance line indicates the full dressed propagator.}
\label{fig:poleterms_and_vertex_correction}
\end{figure}

Let us assume that a scalar resonance (called $\sigma$) is observed in the elastic scattering of (stable) spinless particles described by a field $\phi$. Let us also assume that the mass and the width, as well as the residue of the $\phi\phi$ scattering amplitude at the resonance pole have somehow been extracted to some satisfying accuracy, and that the effects due to inelastic channels are suppressed. We would like to describe this situation with the help of a simple Lagrangian,
\begin{eqnarray}
\mathcal{L}_{\mathrm{basic}} &=& \frac{1}{2}\partial_{\nu}\sigma\partial^{\nu}\sigma - \frac{1}{2}\mu_{\sigma}^{2}\sigma^2 + \frac{1}{2}\partial_{\nu}\phi\,\partial^{\nu}\phi - \frac{1}{2}M_{\phi}^{2}\phi^2 -\frac{g}{2}\sigma\,\phi^2\,,\label{eq:Lsigmamodel}\\ 
\mathcal{L}_{\mathrm{ct}} &=& \frac{\delta Z_{\sigma}}{2}\left(\partial_{\nu}\sigma\partial^{\nu}\sigma -\mu_{\sigma}^{2}\sigma^2\right)- \frac{1}{2}\delta\mu_{\sigma}^{2}\sigma^2 + \frac{\delta Z_{\phi}}{2}\left(\partial_{\nu}\phi\,\partial^{\nu}\phi - M_{\phi}^{2}\phi^2\right) -\frac{1}{2}\delta M_{\phi}^{2}\phi^2 -\frac{\delta g}{2}\sigma\,\phi^2\,.\nonumber
\end{eqnarray}
This model has also been studied e.g. in \cite{Giacosa:2012de}, and in App.~E of \cite{Hyodo:2011qc}, with different methods. As in Sec.~\ref{sec:se} we denote the measured pole position of the $\sigma$ in the complex Mandelstam plane as $s_{\sigma}=\mu_{\sigma}^{2}-i\mu_{\sigma}\gamma_{\sigma}$, and $M_{\phi}$ is the measured mass of the $\phi$ particle. The Lagrangian in the second line contains the counterterms (for the standard procedure in the counterterm approach, see e.g. Chapter 10 in both \cite{Weinberg:1995mt} and \cite{Peskin:1995ev}, or \cite{Collins:1984xc}). The counterterms $\delta Z_{\phi}$ and $\delta M_{\phi}^{2}$ will be adjusted following the usual renormalization conditions, so that the physical mass (the pole of the full propagator) of the $\phi$ equals $M_{\phi}$ to all orders in the loop expansion, while the residue of the propagator is fixed to $1$. We will not discuss the $\phi$ self-energy further. Note that $\delta g$ starts at order $g^3$, while the other counterterms start at $\mathcal{O}(g^2)$. In the following, we will try to fix the remaining counterterms to one-loop accuracy by studying the resonance pole contribution to the $\phi\phi$ scattering amplitude. This contribution is depicted symbolically in Fig.~\ref{fig:poleterms_and_vertex_correction}(a). One finds
\begin{eqnarray}
T^{\mathrm{pole}}_{\phi\phi}(s) &=& -\biggl[1+\frac{\delta g}{g}+\frac{g^{2}}{2}I_{\sigma\phi\phi}(s)\biggr]\frac{g^2}{s-\mu_{\sigma}^{2}-\Pi_{\sigma}(s)}\biggl[1+\frac{\delta g}{g}+\frac{g^{2}}{2}I_{\sigma\phi\phi}(s)\biggr]\,, \\
\Pi_{\sigma}(s) &=& \frac{g^2}{2}I_{\phi\phi}(s)+\delta\mu_{\sigma}^{2}-(s-\mu_{\sigma}^{2})\delta Z_{\sigma}\,.\nonumber
\end{eqnarray}
The loop integrals occuring here are given by
\begin{eqnarray}
I_{\phi\phi}(s\equiv k^2) &:=& \int\frac{d^{d}l}{(2\pi)^{d}}\frac{i}{((k-l)^2-M_{\phi}^{2})(l^2-M_{\phi}^{2})} \nonumber \\
 &=& I_{\phi\phi}(0) -\frac{s}{16\pi^2}\int_{4M_{\phi}^2}^{\infty}ds'\frac{\sigma(s')}{s'(s'-s)} \label{eq:disp1}\\
 &=& I_{\phi\phi}(0)-\frac{1}{8\pi^2}\left(1+\sigma(s)\,\mathrm{artanh}\left(-\frac{1}{\sigma(s)}\right)\right)\,,\quad \sigma(s):=\sqrt{1-\frac{4M_{\phi}^2}{s}}\,,\nonumber\\
I_{\phi\phi}(0) &=& 2\bar{\lambda}+\frac{1}{16\pi^2}\left(1+\log\left(\frac{M_{\phi}^2}{\mu^2}\right)\right)+\mathcal{O}(4-d)\,,\nonumber \\
 \bar\lambda &=& \frac{\mu^{d-4}}{16\pi^{2}}\biggl(\frac{1}{d-4}-\frac{1}{2}[\ln(4\pi)+\Gamma'(1)+1]\biggr) \,,\nonumber \\
I_{\sigma\phi\phi}(s\equiv k^2) &:=& \int\frac{d^{d}l}{(2\pi)^{d}}\frac{i}{((k-l)^2-M_{\phi}^{2})(l^2-M_{\phi}^{2})((k-q-l)^2-\mu_{\sigma}^{2})}\biggr|_{q^2=(k-q)^2=M_{\phi}^{2}} \nonumber \\
 &=& \frac{1}{16\pi^2}\int_{4M_{\phi}^{2}}^{\infty}ds'\,\frac{\log\left(\frac{\mu_{\sigma}^{2}+s'-4M_{\phi}^{2}}{\mu_{\sigma}^{2}}\right)}{\sqrt{s'(s'-4M_{\phi}^{2})}(s'-s)}\,,\label{eq:disp2}
\end{eqnarray}
employing dimensional regularization. From the dispersive representations in Eqs.~(\ref{eq:disp1}) and (\ref{eq:disp2}), the imaginary parts (for real $s$) can be directly read off. The real part of $I_{\phi\phi}(s)$ contains a divergent constant for $d\rightarrow 4$. Real values of $s$ are to be approached from the upper complex plane for $s\in\lbrack 4M_{\phi}^2,\infty\rbrack$ on the physical real axis. Note that, in the application of this appendix, we have to use the expressions for the loop integrals on the unphysical Riemann sheet of the variable $s$, which can be obtained by analytic continuation in $s$.\\

To avoid that the real part of the pole position is shifted from its value $\mu_{\sigma}^{2}$, we have to fix $\delta\mu_{\sigma}^{2}=-\frac{g^2}{2}\mathrm{Re}\,I_{\phi\phi}(s_{\sigma})$. To assure that the imaginary part of the pole position equals $-\mu_{\sigma}\gamma_{\sigma}$, we need
\begin{equation}
\mu_{\sigma}\gamma_{\sigma}(1+\delta Z_{\sigma})=-\frac{g^2}{2}\mathrm{Im}\,I_{\phi\phi}(s_{\sigma})\quad \Rightarrow\quad \gamma_{\sigma} = -\frac{g^2}{2\mu_{\sigma}}\mathrm{Im}\,I_{\phi\phi}(s_{\sigma}) +\mathcal{O}(g^4)\,,
\end{equation}
which fixes the value for $g$ in our theory. Then, close to the pole,
\begin{eqnarray}
T^{\mathrm{pole}}_{\phi\phi}(s) &\rightarrow& -\frac{g^2}{s-s_{\sigma}}\frac{\left[1+(\delta g/g)+(g^{2}/2)I_{\sigma\phi\phi}(s_{\sigma})\right]^2}{\left[1+\delta Z_{\sigma}-(g^{2}/2)I_{\phi\phi}'(s_{\sigma})\right]} \nonumber \\
 &\approx& -\frac{g^2}{s-s_{\sigma}}\biggl[1+\left(\frac{2\delta g}{g}-\delta Z_{\sigma}\right)+g^2\left(\mathrm{Re}\,I_{\sigma\phi\phi}(s_{\sigma})+\frac{1}{2}\mathrm{Re}\,I_{\phi\phi}'(s_{\sigma})\right) \nonumber \\ &+&\quad ig^2\left(\mathrm{Im}\,I_{\sigma\phi\phi}(s_{\sigma})+\frac{1}{2}\mathrm{Im}\,I_{\phi\phi}'(s_{\sigma})\right)\biggr]\,,
\end{eqnarray}
where in the second line we have neglected terms of two-loop order. We see that only a combination of wave-function and coupling counterterms can be fixed from the real part of the residue at the pole, and that the imaginary part of the residue is fixed by $g$ and $\mu_{\sigma}$ at one-loop order. The usual choice for $\delta Z_{\sigma}$ would be $(g^2/2)\mathrm{Re}\,I_{\phi\phi}'(s_{\sigma})$, in which case the {\em real part}\, of the residue of the resonance propagator is fixed to $1$. This choice was adopted e.g. in \cite{Bruns:2013tja,Denner:1991kt,Klingl:1996by} (compare also \cite{Djukanovic:2009zn} for the ``complex mass scheme'').
More generally, the residue of the scattering amplitude at the resonance pole is determined by corrections to wave-function renormalization, and by vertex corrections. The latter are essentially given by the scattering amplitude on the second Riemann sheet, {\em with the pole term subtracted}, evaluated at the resonance pole. The $\sigma$ exchange in Fig.~\ref{fig:poleterms_and_vertex_correction}(b) is a part of this subtracted amplitude. Intuitively it should be the vertex correction due to this subtracted scattering amplitude that is strongly related to the concept of ``compositeness'' of the resonance \cite{Hyodo:2011qc,Sekihara:2014kya} (besides the resonance location), since apparently it determines the ``overlap'' matrix element between the ``resonance state'' and a state of two particles interacting via $t-$ and $u-$ channel exchanges (so we should expect resonances with a noteable composite two-particle component to be sensitive to $t-$ and $u-$ channel dynamics). The subtracted scattering amplitude in question could be extracted employing unitarity and analyticity: it is well-known \cite{Basdevant:1973ru,Caprini:2005zr} that the partial-wave S-matrix elements for elastic scattering, $S_{\ell}=1+2i\sigma(s)t_{\ell}(s)$ on the first ($I$) and second ($II$) sheet are related by $S_{\ell}^{II}=1/S_{\ell}^{I}$ (note that $\sigma(s)$ in Eq.~(\ref{eq:disp1}) also has a branch point at the threshold). So, numerically, the zeros $s_{0}$ on the first (physical) sheet agree with the resonance pole positions $s_{\sigma}$ on the second sheet. Expanding $S_{\ell}^{I}(s)=0+(s-s_{0})a_{1}+\frac{1}{2}(s-s_{0})^2a_{2}+\ldots\,$, we find
\begin{equation}
S_{\ell}^{II}(s\rightarrow s_{\sigma}) = \frac{a_{1}^{-1}}{s-s_{\sigma}}-\frac{a_{2}}{2a_{1}^{2}}\,,
\end{equation}
which gives us the subtracted S-matrix element at the resonance position on the second sheet. This is an interesting result - however, the page ends here.

\end{appendix}

\end{document}